\documentclass[aps,twocolumn,showpacs,superscriptaddress]{revtex4}
\usepackage{epsf}
\usepackage{multirow}

\begin{document}

\title{Search on a Hypercubic Lattice using a Quantum Random Walk: I. $d>2$}
\author{Apoorva Patel}
\email{adpatel@cts.iisc.ernet.in}
\affiliation{Centre for High Energy Physics,
             Indian Institute of Science, Bangalore-560012, India}
\affiliation{Supercomputer Education and Research Centre,
             Indian Institute of Science, Bangalore-560012, India}
\author{Md. Aminoor Rahaman}
\email{aminoorrahaman@yahoo.com}
\affiliation{Supercomputer Education and Research Centre,
             Indian Institute of Science, Bangalore-560012, India}
\date{\today}

\begin{abstract}
\noindent
Random walks describe diffusion processes, where movement at every time step
is restricted to only the neighbouring locations. We construct a quantum
random walk algorithm, based on discretisation of the Dirac evolution operator
inspired by staggered lattice fermions. We use it to investigate the spatial
search problem, i.e. find a marked vertex on a $d$-dimensional hypercubic
lattice. The restriction on movement hardly matters for $d>2$, and scaling
behaviour close to Grover's optimal algorithm (which has no restriction on
movement) can be achieved. Using numerical simulations, we optimise the
proportionality constants of the scaling behaviour, and demonstrate the
approach to that for Grover's algorithm (equivalent to the mean field theory
or the $d\rightarrow\infty$ limit). In particular, the scaling behaviour for
$d=3$ is only about 25\% higher than the optimal $d\rightarrow\infty$ value.
\end{abstract}
\maketitle

\section{Spatial Search using the\break Dirac Operator}

The spatial search problem is to find a marked object from an unsorted
database of size $N$ spread over distinct locations. Its characteristic
is that one can proceed from any location to only its neighbours, while
inspecting the objects. The problem is conventionally represented by a
graph, with the vertices denoting the locations of objects and the edges
labeling the connectivity of neighbours. Classical algorithms for this
problem are $O(N)$, since they can do no better than inspect one location
after another until reaching the marked object. On the other hand,
quantum algorithms can do better in search problems by working with a
superposition of states, Grover's algorithm being the prime example
\cite{grover}. The spatial search problem has been a focus of investigation
in recent years using different quantum algorithmic techniques, both
analytical and numerical, and in a variety of spatial geometries ranging
from a single hypercube to regular lattices
\cite{hypsrch1,gridsrch1,gridsrch2,tulsi,hypsrch2,hexsrch}.
These studies have mainly looked at the asymptotic scaling forms of the
algorithms, and have not varied the database size $N$ and its dimension
$d$ independently. In this work, we study the specific case of searching
for a marked vertex on a $d$-dimensional hypercubic lattice with $N=L^d$
vertices. We let $N$ and $d$ be independent, as well as determine the
scaling prefactors, and thereby develop a broad picture of how the
dimension of the database (or the connectivity of the graph) influences
the spatial search problem.

The quantum algorithmic strategy for spatial search is to construct a
Hamiltonian evolution, where the kinetic part of the Hamiltonian diffuses
the amplitude distribution all over the lattice and the potential part of
the Hamiltonian attracts the amplitude distribution toward the marked
vertex \cite{grover_strategy}. The optimization criterion for the algorithm
is to concentrate the amplitude distribution toward the marked vertex as
quickly as possible. Grover constructed the optimal global diffusion
operator, but it requires movement from any vertex to any other vertex in
just one step. That corresponds to a fully connected graph, or equivalently
the mean field theory limit. When movements are restricted to be local
(i.e. one can go from a vertex to only its neighbours in one step), the
search algorithm slows down. It is then worthwhile to redesign the search
algorithm by understanding the extent of slow down due to local movements,
and how it depends on the spatial arrangement of the database. That is the
question we address quantitatively in this article. As discussed in what
follows, for the best spatial search algorithms, the slow down is in the
scaling prefactor for $d>2$ and in the scaling form for $d\le 2$.

A diffusion process can be modeled either using continuous time and a first
order time derivative, or using discrete time and a single time step evolution.
In its discrete version, it is generically described as a random walk,
whereby a probability distribution evolves in a non-deterministic manner
at every time step. Such random walks have been used to tackle a wide
range of graph theory problems, usually with a local and translationally
symmetric evolution rule. We use a quantum version of this process, i.e.
quantum random walks \cite{qrw}. It provides a unitary evolution of the
quantum amplitude distribution, such that the amplitude at each vertex
gets redistributed over itself and its neighbours at every time step.
It is worth noting that quantum random walks are deterministic, unlike
classical random walks, with quantum superposition allowing simultaneous
exploration of multiple possibilities. Several quantum algorithms have
used them as important ingredients, and an introductory overview can be
found in Ref.\cite{qrwrev}.

On a periodic lattice with translational symmetry, spatial propagation modes
are characterized by their wave vectors $\vec{k}$. Quantum diffusion depends
on the energy of these modes according to $U(\vec{k},t)=\exp(-iE(\vec{k})t)$.
The lowest energy mode, $\vec{k}=0$, corresponding to a uniform distribution,
is an eigenstate of the diffusion operator and does not propagate.
The slowest propagating modes are the ones with the smallest nonzero
$|\vec{k}|$. The commonly used diffusion operator is the Laplacian, with
$E(\vec{k}) \propto|\vec{k}|^2$. The alternative Dirac operator, available
in a quantum setting, provides a faster diffusion of the slowest modes with
$E(\vec{k})\propto|\vec{k}|$. Its quadratically faster spread makes it better
suited to quantum spatial search algorithms \cite{gridsrch1,gridsrch2}.

An automatic consequence of the Dirac operator is the appearance of an internal
degree of freedom corresponding to spin, whereby the quantum state becomes
a multi-component spinor. These spinor components can guide the diffusion
process, and be interpreted as the states of a coin \cite{gridsrch1,gridsrch2}.
While this is the only possibility for the continuum theory, another option
is available for a lattice theory, i.e. the staggered fermion formalism
\cite{staggered}. In this approach, the spinor degrees of freedom are spread
in coordinate space over an elementary hypercube, instead of being in an
internal space. We follow it to construct a quantum search algorithm on a
hypercubic lattice, reducing the total Hilbert space dimension by $2^d$ and
eliminating the coin toss instruction.

\subsection{Bipartite Decomposition}

The free particle Dirac Hamiltonian in $d$ dimensions is
\begin{equation}
H_{\rm free} = -i\vec{\alpha}\cdot\vec{\nabla} + \beta m ~.
\label{diracham}
\end{equation}
On a hypercubic lattice, with the simplest discretization of the gradient
operator,
\begin{equation}
\nabla_n f(\vec{x}) = {1 \over 2}[ f(\vec{x}+\hat{n}) - f(\vec{x}-\hat{n}) ] ~,
\end{equation}
and a convenient choice of basis, the anticommuting matrices
$\vec{\alpha},\beta$ are spin-diagonalized to location dependent signs:
\begin{eqnarray}
\psi \rightarrow T \psi ,~ H \rightarrow T H T^{\dagger} ,&&
T = \alpha_{d}^{x_{d}} \alpha_{d-1}^{x_{d-1}}
  \cdots \alpha_{2}^{x_{2}} \alpha_{1}^{x_{1}} ,\\
\alpha_n \rightarrow \prod_{j=1}^{n-1} (-1)^{x_j} ~,&&
\beta    \rightarrow \prod_{j=1}^{d} (-1)^{x_j} \beta ~.
\label{pmsigns}
\end{eqnarray}
With this choice, translational invariance holds in steps of 2 (instead of 1),
and the squared Hamiltonian,
\begin{equation}
H_{\rm free}^2 = -\vec{\nabla}\cdot\vec{\nabla} + m^2 ~,
\label{dirachamsq}
\end{equation}
has eigenvalues $\sum_n \sin^2(k_n)+m^2$ on a hypercubic lattice.
Inclusion of a mass term slows down the diffusion process, but it also
regulates the $\vec{k}=0$ mode. In the present article, we set $m=0$;
in an accompanying article \cite{deq2search}, we investigate the
advantage a nonzero mass offers when the problem has an infrared divergence.

\begin{figure}[b]
\epsfxsize=8cm
\centerline{\epsfbox{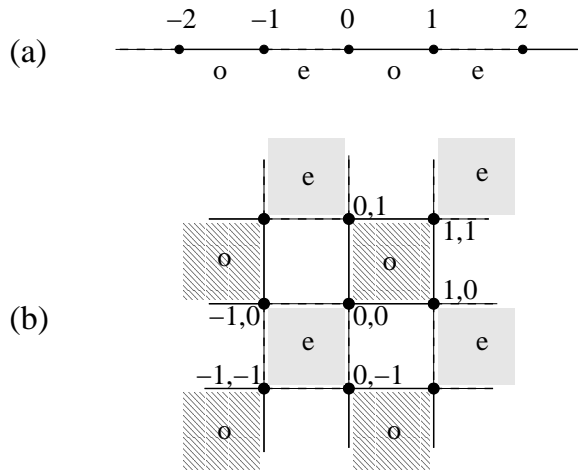}}
\caption{Bipartite decomposition of the hypercubic lattice
in to odd and even parts: (a) for $d=1$, (b) for $d=2$.}
\end{figure}

Even when the Hamiltonian $H$ is local, the discrete time evolution
operator $U=\exp(-iH\tau)$ may connect arbitrarily separated vertices.
To make $U$ also local, we split $H$ in to block-diagonal Hermitian parts
and then exponentiate each part separately \cite{qwalk1}. For a bipartite
lattice, partitioning of $H$ in to two parts (which we label ``odd'' and
``even'') is sufficient for this purpose \cite{spinorsize}:
\begin{equation}
H_{\rm free} = H_o + H_e ~.
\end{equation}
This partition is illustrated in Fig.1 for $d=1$ and $d=2$. Each part
contains all the vertices, but only half of the links attached to each vertex.
Each link is associated with a term in $H_{\rm free}$ providing propagation
along it, and appears in only one of the two parts. $H$ thus gets divided
in to a set of non-overlapping blocks of size $2^d \times 2^d$ (each block
corresponds to an elementary hypercube on the lattice) that can be exactly
exponentiated. The amplitude distribution then evolves according to
\begin{equation}
\psi(\vec{x};t) = W^t \psi(\vec{x};0) ~,~~
W = U_e U_o = e^{-iH_e\tau} e^{-iH_o\tau} ~.
\end{equation}
Each block of the unitary matrices $U_{o(e)}$ mixes amplitudes of vertices
belonging to a single elementary hypercube, and the amplitude distribution
spreads in time because the two alternating matrices do not commute. Note
that $W$ does not perform evolution according to $H_{\rm free}$ exactly.
Instead, $W=\exp(-iH_{\rm free}\tau) + O(\tau^2)$. Still, the truncation is
such that $W$ is exactly unitary, i.e. $W=\exp(-i\tilde{H}\tau)$ for some
$\tilde{H}$.

We adopt the convention where the walk operator $W$ is real. For $d=1$,
we set
\begin{eqnarray}
H_o |x\rangle &=& -{i \over 2}\Big[ (-1)^x |x+(-1)^x\rangle \Big] ~, \\
H_e |x\rangle &=&  {i \over 2}\Big[ (-1)^x |x-(-1)^x\rangle \Big] ~.
\end{eqnarray}
The resultant Hamiltonian blocks are $H_o^B=-\sigma_2/2$, and
$H_e^B=\sigma_2/2$ when operating on the block with coordinates flipped
in sign. With $(H_o^B)^2 = (H_e^B)^2 = {1 \over 4}I$,
\begin{equation}
U_{o(e)} = cI - 2is H_{o(e)} ~,~~ |c|^2 + |s|^2 = 1 ~,
\end{equation}
where $s=\sin(\tau/2)$ is a parameter that can be tuned.

These expressions can be rearranged to separate the translationally
invariant part of the walk from the location dependent signs. Let the
amplitude distribution in a two spinor (or coin) component notation be
\begin{equation}
\Psi(X,t) \equiv \left(\matrix{ \psi(2x,t) \cr \psi(2x-1,t) \cr}\right) ~.
\end{equation}
Then the walk operator becomes
\begin{eqnarray}
W &=& \tilde{U}C ~,~~
C = U_{o}^{B} = \left(\matrix{ c & s \cr -s & c \cr}\right) ~, \\
\tilde{U}|X\rangle &=& c|X\rangle
                    -  s\sum_{\pm} (\pm\sigma_{\pm})|X\pm1\rangle ~.
\end{eqnarray}
The operator $C=U_{o}^{B}$ just mixes the components of $\Psi$ within
the block. The first term of the operator $\tilde{U}$ is stationary,
while the other moves the amplitude to neighbouring blocks. The movement
is accompanied by chirality flip, a characteristic feature of the Dirac
operator, because of the raising and lowering Pauli operators
$\sigma_{\pm}=(\sigma_{1}\pm i\sigma_{2})/2$.

In $d$ dimensions, the preceding structure generalises to $2^d\times2^d$
Hamiltonian blocks that can be expressed in terms of tensor products
of Pauli matrices \cite{qwalk2}. When operating on hypercubes with
coordinate labels $\{0,1\}^{\otimes d}$,
\begin{equation}
H_o^B = -{1\over2} \sum_{j=1}^d
      I^{\otimes (d-j)} \otimes\sigma_2\otimes \sigma_3^{\otimes (j-1)} ~,
\label{hoblock}
\end{equation}
and $H_e^B = -H_o^B$ when operating on a hypercube with all coordinates
flipped in sign. Note that Eq.(\ref{hoblock}) is the sum of $d$ terms,
which mutually anticommute and whose squares are proportional to identity.
This is the characteristic signature of the Dirac Hamiltonian,
Eq.(\ref{diracham}). The block-diagonal matrices satisfy $H_o^2 = H_e^2 =
{d \over 4}I$, and exponentiate to 
\begin{equation}
U_{o(e)} = cI - is{2\over\sqrt{d}}H_{o(e)} ~,~~ |c|^2 + |s|^2 = 1 ~.
\end{equation}
The parameter $s=\sin(\sqrt{d}\tau/2)$ can be optimised, to achieve the
fastest diffusion across the lattice. Note that the $d$-dimensional walk
is not the tensor production of $d$ one-dimensional walks.

Again separating the degrees of freedom within each local hypercube as
$\vec{x} \equiv \vec{X}\otimes$\{0,-1\}$^{\otimes d}$,
the location dependent signs can be rewritten as operators acting on the
spinor (or coin) degrees of freedom. The walk operator takes the form:
$W=\tilde{U}C$, $C=U_{o}^{B}$, and
\begin{eqnarray}
\tilde{U}|\vec{X}\rangle &=& c|\vec{X}\rangle \\
         &-& \frac{s}{\sqrt{d}} \sum_{\pm}\sum_{j=1}^{d}
         (I^{\otimes(j-1)} \otimes (\pm\sigma_{\pm}) \otimes
         \sigma_{3}^{\otimes(d-j)}) |\vec{X}\pm\hat{j}\rangle ~. \nonumber
\end{eqnarray}

\subsection{Search Algorithm}

To search for a marked vertex, say the origin, we attract the quantum
random walk toward it by adding a potential to the free Hamiltonian,
\begin{equation}
V = \cases{ \beta V_0 ~\delta_{\vec{x},0} ~~ ({\rm gravitational}) ~, \cr
                  V_0 ~\delta_{\vec{x},0} ~~ ({\rm electromagnetic}) ~.}
\end{equation}
Exponentiation of this potential produces a phase change for the amplitude
at the marked vertex. For the steepest attraction of the quantum random walk
toward the marked vertex, it is optimal to choose $V_0$ so as to make the
corresponding evolution phase maximally different from $1$, i.e.
$e^{-iV_0\tau}=-1$. The sign provided by $\beta$ does not matter in this case,
and the evolution phase becomes the binary oracle,
\begin{equation}
R = I - 2 |\vec{0}\rangle\langle\vec{0}| ~.
\end{equation}
The search algorithm alternates between diffusion and oracle
operators---a discrete version of Trotter's formula involving
$H_o$, $H_e$ and $V$---yielding the evolution
\begin{equation}
\psi(\vec{x};t_1,t_2) = [W^{t_1} R]^{t_2} \psi(\vec{x};0,0) ~.
\label{evolsearch}
\end{equation}
Here $t_2$ is the number of search oracle queries, and $t_1$ is the number
of walk steps between queries. Both should be minimised to find the the
quickest solution to the search problem.

Since this algorithm iterates a unitary operator on the initial state,
it produces periodic results. As a function of the number of iterations,
the probability of being at the marked vertex first increases, reaches
a peak value $P$ after $t_2$ search oracle queries, then decreases 
toward zero, and thereafter keeps on repeating this cycle. Grover's
algorithm is designed to evolve the quantum state in a two-dimensional
Hilbert space (spanned by the initial and marked states), and is able
to reach $P=1$. That does not hold for spatial search, and the resultant
values of $P$ are less than $1$. The remedy is to augment the algorithm
by the amplitude amplification procedure \cite{BHMT}, to find the marked
vertex with $\Theta(1)$ probability. The complexity of the algorithm is
then characterised by the effective number of search oracle queries,
$t_2/\sqrt{P}$.

\subsection{Complexity Bounds}

Spatial search obeys two simple lower bounds. One arises from the fact
that while the marked vertex could be anywhere on the lattice, one step
of the local walk can move from any vertex to only its nearest neighbours.
Since the marked vertex cannot be located without reaching it, the worst
case scenario on a hypercubic lattice requires $\Omega(dL)$ steps.
This bound is the strongest in one dimension, where quantum search is
unable to improve on the $O(N)$ classical search.

The other bound follows from the fact that spatial search cannot outperform
Grover's optimal algorithm, and so it must require $\Omega(\sqrt{N})$ oracle
queries. That is independent of $d$, and stronger than the first bound for
$d>2$. As a matter of fact, the first bound weakens with increasing $d$,
as the number of neighbours of a vertex increase and steps required to go
from one corner of the lattice to another decrease. In analogy with mean
field theory behaviour of problems in statistical mechanics, we can thus
expect that in the $d\rightarrow\infty$ limit the locality restriction on
the walk would become irrelevant and the complexity of spatial search would
approach that of Grover's algorithm. (Note that the maximum value of $d$ is
$\log_2 N$ for finite $N$.) Our results in this work demonstrate that not
only is the $\Omega(\sqrt{N})$ complexity scaling achievable for $d>2$,
but one can get pretty close to the asymptotic prefactor $\pi/4$ as well.

Combined together, the two bounds make the complexity of spatial search
$\Omega(d N^{1/d},\sqrt{N})$. The two bounds are of the same magnitude,
$\Omega(\sqrt{N})$, for the critical case of $d=2$. It is a familiar
occurrence in statistical mechanics of critical phenomena that an interplay
of different dynamical features produces logarithmic correction factors
in critical dimensions due to infrared divergences. The same seems to be
true for spatial search in $d=2$, where algorithms are slowed down by
extra logarithmic factors \cite{gridsrch1,gridsrch2,tulsi}. It is an open
question to design algorithms, perhaps using additional parameters, that
suppress the logarithmic factors as much as possible. We look at the
special $d=2$ case in an accompanying article.

The way the lower bounds arise in spatial search also illustrates an
interplay of two distinct physical principles, special relativity (or
no faster than light signaling) and unitarity. It is well-known that
the two are compatible, but just barely so, in the sense that physical
theories that are unitary but not relativistic or relativistic but not
unitary exist. Furthermore, the best algorithms should arise in a
framework that respects both the principles, i.e. relativistic quantum
mechanics. In spatial search, locality of the quantum walk and the
square-root speed up produced by the Dirac equation (through change in
the dispersion relation) follow from special relativity. On the other
hand, optimality of square-root speed up provided by Grover's algorithm
is a consequence of unitarity \cite{BBBV,zalka}. Why these two distinct
physical principles lead to the same scaling constraint is not understood,
and is a really interesting question to explore. At present, however,
we have nothing more to add.

We point out that preparation of the unbiased initial state for the search
problem, i.e. the translationally invariant uniform superposition state
$|s\rangle=\sum_x|\vec{x}\rangle/\sqrt{N}$, is not at all difficult using
local directed walk steps. For instance, one can start at the origin,
step by step transfer the amplitude to the next vertex along an axis,
and achieve an amplitude $L^{-1/2}$ at all the vertices on the axis after
$L$ steps. Repeating the procedure for each remaining coordinate direction
produces the state $|s\rangle$ after $dL$ steps in total. Clearly, this
preparation does not add to the complexity of the algorithm.

\section{Optimisation of Parameters}

The fastest search amounts to finding the shortest unitary evolution path
between the initial state $|s\rangle$ and the marked state $|\vec{0}\rangle$.
This path is a geodesic arc on the unitary sphere from $|s\rangle$ to
$|\vec{0}\rangle$, and Grover's optimal algorithm strides along this path
perfectly. In our algorithm, Eq.(\ref{evolsearch}), we have replaced
Grover's optimal diffusion operator, $G=2|s\rangle\langle s|-1$, with the
walk operator $W^{t_1}$. So to optimise our algorithm, we need to tune
the parameters appearing in $W^{t_1}$, i.e. $s$ (or $c$) and $t_1$,
such that $W^{t_1}R$ approximates a rotation in the two-dimensional
$|s\rangle$-$|\vec{0}\rangle$ subspace by the largest possible angle.

The operator $G$ has $|s\rangle$ as an eigenstate with eigenvalue $1$.
All other states orthogonal to $|s\rangle$, in particular the combination
$|s_\perp\rangle = (|s\rangle-\sqrt{N}|\vec{0}\rangle)/\sqrt{N-1}$
in the $|s\rangle$-$|\vec{0}\rangle$ subspace, are its eigenstates with
eigenvalue $-1$. The walk operator $W=U_{e}U_{o}$ also has $|s\rangle$
as an eigenstate with eigenvalue $1$. However, its other eigenvalues are
spread around the unit circle, and evolution under $W^{t_1}R$ does not
remain fully confined to the $|s\rangle$-$|\vec{0}\rangle$ subspace. If the
outward diffusion is not controlled, the probability of remaining in the
$|s\rangle$-$|\vec{0}\rangle$ subspace would decay exponentially with the
number of iterations. To reach the marked state with a constant probability,
therefore, parameters must be tuned so that part of the outward diffusion
subsequently returns to the $|s\rangle$-$|\vec{0}\rangle$ subspace.

\subsection{Analytical Criteria}

The optimisation condition that makes $W^{t_1}$ as close an approximation
to $G$ as possible can be formulated in several different ways. Though they
all lead to the same result, describing them separately is instructive.

(1) Maximise the overlap of $W^{t_1}$ and $G$, i.e.
\begin{equation}
Tr(W^{t_1}G)  =  Tr(2|s\rangle\langle s| - W^{t_1})
              = 2 - \sum_i \langle i|W^{t_1}|i \rangle ~.
\end{equation}
$\langle i|W^{t_1}|i \rangle$ is the amplitude for the random walk to
return to the starting state $|i\rangle$ after $t_1$ time steps. Since $W$
is translationally invariant along each coordinate direction in steps of 2,
we need to evaluate $\langle i|W^{t_1}|i \rangle$ only for states $|i\rangle$
in an elementary hypercube. Now, any closed path on a hypercubic lattice
takes an even number of steps along each coordinate direction, and all
location dependent propagation signs (cf. Eq.(\ref{pmsigns})) get squared
to $1$ in the process. As a result, $\langle i|W^{t_1}|i \rangle$ is
independent of $|i\rangle$, and the optimisation condition becomes the
minimisation of $A(t_1)\equiv\langle \vec{0}|W^{t_1}|\vec{0} \rangle$.

(2) Make $|s_\perp\rangle$ approximate an eigenvector of $W^{t_1}$,
with an eigenvalue approaching $-1$, i.e. minimise
\begin{equation}
\langle s_\perp|W^{t_1}|s_\perp \rangle
= {1 \over N-1}\big(-1 + N A(t_1)\big) ~,
\end{equation}
which happens to be the same condition as given earlier. In addition,
unitarity of the evolution operator implies
\begin{equation}
|\langle s_\perp|W^{t_1}|s_\perp \rangle| \leq 1
  ~\Longrightarrow~ -1 + {\textstyle 2 \over N} \leq A(t_1) \leq 1 ~.
\end{equation}

(3) Maximise the rotation provided by the operator $W^{t_1}R$ along the
$|s\rangle$-$|\vec{0}\rangle$ subspace. This rotation is given by the
projection
\begin{eqnarray}
&& {\rm Proj}_{s,s_\perp}[W^{t_1}R] = \pmatrix{
\langle s      |W^{t_1}R|s\rangle & \langle s      |W^{t_1}R|s_\perp\rangle \cr
\langle s_\perp|W^{t_1}R|s\rangle & \langle s_\perp|W^{t_1}R|s_\perp\rangle \cr}
\nonumber\\
&& = \pmatrix{
1 - {2\over N} & {2\over N}\sqrt{N-1} \cr
{2\over\sqrt{N-1}}(A(t_1)-{1\over N}) & (1-{2\over N}){1-NA(t_1)\over N-1} \cr} \label{projWR}\\ 
&& = \pmatrix{ 1 & 0 \cr
               0 & {1-NA(t_1) \over N-1} \cr }
     \pmatrix{ 1-{2\over N} & {2\over N}\sqrt{N-1} \cr
              -{2\over N}\sqrt{N-1} & 1-{2\over N} \cr} ~. \nonumber
\end{eqnarray}
The factor on the right in Eq.(\ref{projWR}) is precisely the rotation matrix
produced by the operator $GR$ in Grover's algorithm, i.e. rotation by an
angle $2\sin^{-1}(1/\sqrt{N})$ in the $|s\rangle$-$|\vec{0}\rangle$ subspace.
So the total projection can be viewed as first applying an iteration of
Grover's algorithm to the state, and then shrinking the $|s_\perp\rangle$
component by a factor $(1-NA(t_1))/(N-1)$. The shrinking disappears and
the projection tends to an exact orthogonal transformation, as $A(t_1)$
approaches its lower bound $(2-N)/N$. It also follows that when the
shrinking is controlled, cutting down outward diffusion from the
$|s\rangle$-$|\vec{0}\rangle$ subspace, the algorithm will have the same
$O(\sqrt{N})$ scaling as in case of Grover's algorithm.

The minimisation condition for $A(t_1)$ is obtained here only using
the translational invariance of the walk operator $W$, together with
$W|s\rangle = |s\rangle$ and $\langle s|W = \langle s|$. It is easy to
perform this optimisation numerically, as a function of the parameters
$s$ and $t_1$, and our results are described later. Before that, an
understanding of the eigenspectrum of $W$ gives us some idea regarding
what to expect.

Since $H_{o}^{2} = H_{e}^{2} = \frac{d}{4}I$, eigenvalues of $H_{o}$ and
$H_{e}$ are $\pm\sqrt{d}/2$. Moreover, $H_{o}$ and $H_{e}$ are traceless,
so that half the eigenvalues are $\sqrt{d}/2$ and the other half
$-\sqrt{d}/2$. Consequently, eigenvalues of $U_{o}$ and $U_{e}$ are
$c\pm is=\exp(\pm i\sqrt{d}\tau/2)$, equally divided among complex
conjugate pairs. Furthermore, $H_{o(e)}$ and $U_{o(e)}$ have the same
eigenvectors.

Since $U_o$ and $U_e$ do not commute, we do not have a general solution
for the eigenspectrum of $W$. Still it is straightforward to work it out
in the continuum time limit, $\tau\rightarrow0$. $H_{\rm free}$ is
diagonalised by a Fourier transform,
\begin{eqnarray}
H_{\rm free}^{2}|\vec{x}\rangle &=& -\frac{1}{4}\sum_{n=1}^{d} \Big[
  |\vec{x}+2\hat{n}\rangle-2|\vec{x}\rangle+|\vec{x}-2\hat{n}\rangle
  \Big] \nonumber\\
  &\longrightarrow& \sum_{n=1}^{d} \sin^{2}(k_{n}) |\vec{k}\rangle ~,
\end{eqnarray}
with a uniform distribution in the Brillouin zone. Up to $O(\tau^{2})$,
therefore, the eigenvalues of $W (\approx e^{-iH_{\rm free}\tau})$ are
$\exp(\pm i\tau\sqrt{\sum_{j}\sin^{2}k_{j}})$, with the eigenphases in
the interval $[-\tau\sqrt{d},\tau\sqrt{d}]$.
The average value of $H_{\rm free}^{2}$ is $d/2$, and the corresponding
eigenvalues of $W$ are $\exp(\pm i\tau\sqrt{d/2})$. We find that this
average value plays an important role in determination of the optimal
parameters.

\subsection{Numerical Optimisation}

The considerations of the previous subsection focused on what happens
during a single iteration of the algorithm. Over many iterations,
there is outward diffusion of the amplitude distribution from the
$|s\rangle$-$|\vec{0}\rangle$ subspace, and subsequent partial return.
That plays an important role in the overall success of the algorithm,
but we have not been able to estimate it analytically. To study the
performance of the whole algorithm, therefore, we numerically tune the
parameters $s$ and $t_1$ to (a) maximise the probability of reaching
the marked vertex, and (b) minimise the number of search oracle queries
required to reach that stage. The ensuing results of our optimisation
of $P$, $t_2$ and $A(t_1)$ are described in what follows.

We first looked at how the probability distribution evolves as a function
of time during the search process. To illustrate our results, we take
$s=c=1/\sqrt{2}$ that corresponds to maximal mixing between even and odd
vertices, and analyse the evolution as a function of $t_1$ and $t_2$ on
a $32^3$ lattice. As displayed in Fig.2, the probability at the marked
vertex undergoes a cyclic evolution pattern as a function of time $t_2$.
We observe that the evolution is essentially sinusoidal for $t_1=1,2,3$,
and persists for more than 30 cycles without any visible deviation. This
behaviour implies that a constant fraction of the quantum state remains
in the $|s\rangle$-$|\vec{0}\rangle$ subspace, and the operator $W^{t_1}R$
rotates it at a uniform rate within the subspace, fully matching the exact
solution of Grover's algorithm. The difference among $t_1=1,2,3$ is that
the cycle time and the peak probability vary, meaning that the fraction
of the quantum state present in the $|s\rangle$-$|\vec{0}\rangle$ subspace
depends on $t_1$.

For $t_1>3$, we find that the evolution rapidly loses its sinusoidal
nature and the peak probability plummets. The change in the pattern is
so drastic that the algorithm no longer performs a reasonable search.
This behaviour implies that most of the quantum state has moved out of the
$|s\rangle$-$|\vec{0}\rangle$ subspace and only a negligible component is
left behind. An appropriate choice of $t_1$ is thus crucial for constructing
a successful search algorithm.

\begin{figure}[b]
\epsfxsize=9cm
\centerline{\epsfbox{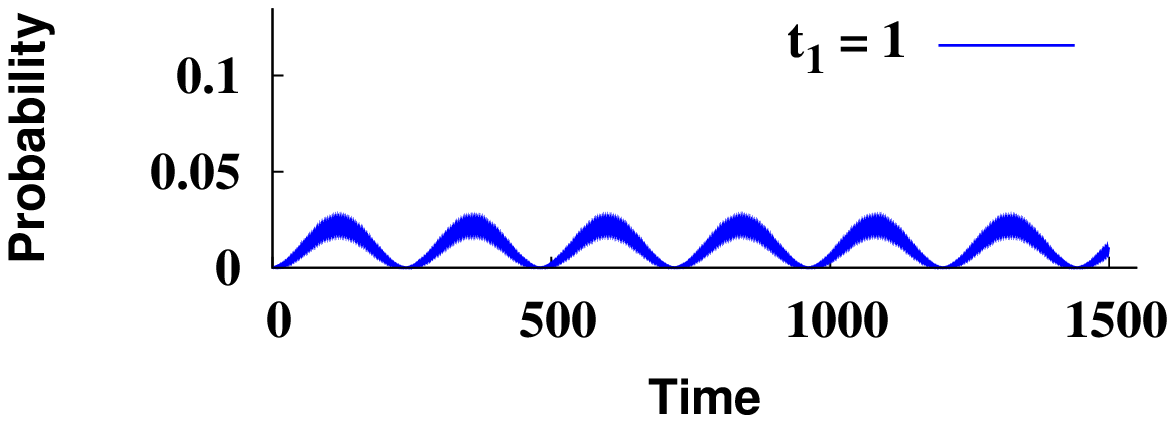}}
\epsfxsize=9cm
\centerline{\epsfbox{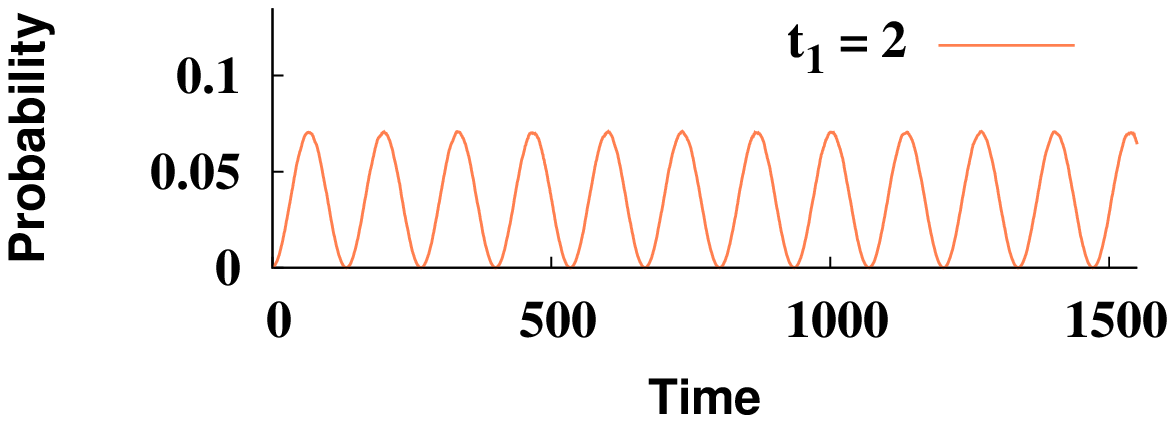}}
\epsfxsize=9cm
\centerline{\epsfbox{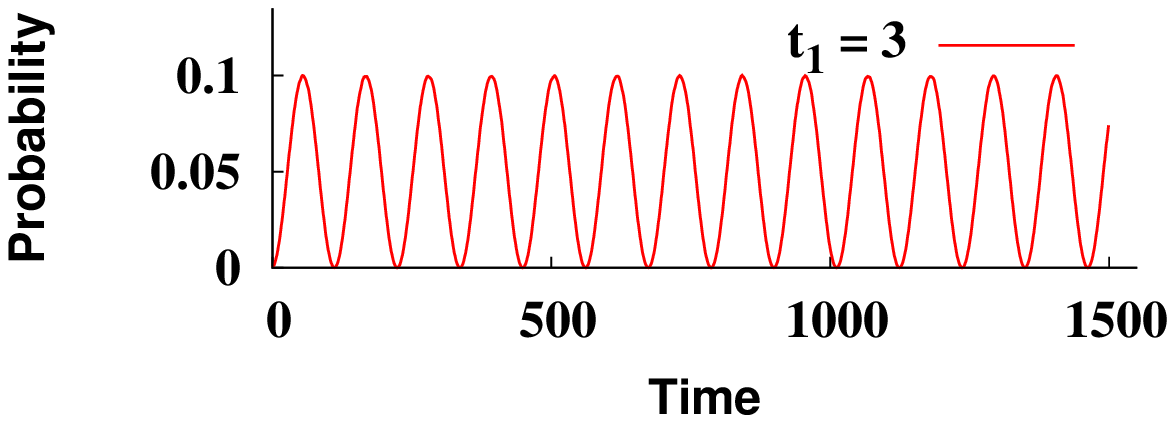}}
\epsfxsize=9cm
\centerline{\epsfbox{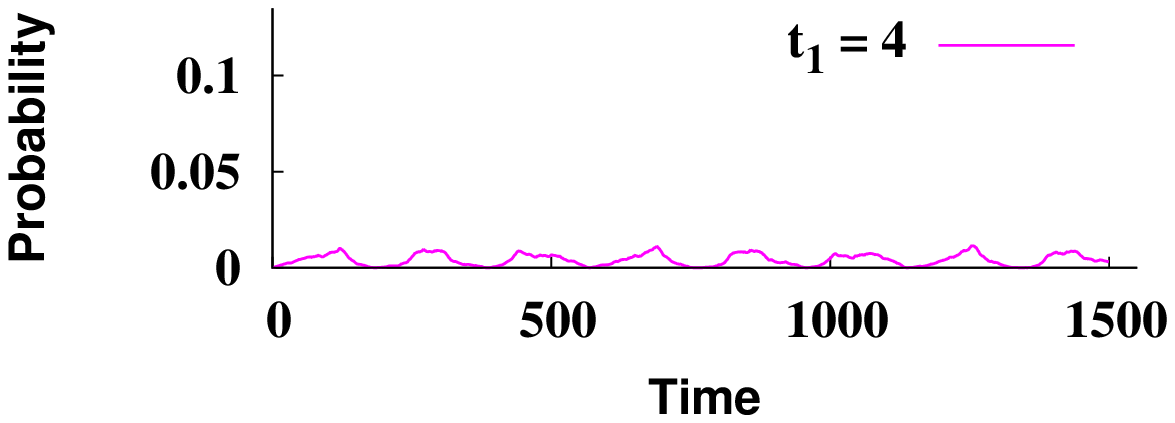}}
\caption{(Color online) Time evolution of the probability at the marked
vertex for the spatial search problem on a $32^3$ lattice. $s=1/\sqrt{2}$
and $t_1$ increases from $1$ (top) to $4$ (bottom).}
\end{figure}

We also show in Fig.3 what the probability distribution looks like at the
instance when the probability at the marked vertex attains its peak value.
We note that the distribution is strongly peaked around the marked vertex
in an essentially uniform background. The peak is sharper when the peak
probability is larger, and it almost disappears for $t_1>3$. This pattern
suggests that the part of the quantum state that diffuses outside the
$|s\rangle$-$|\vec{0}\rangle$ subspace is almost featureless.

\begin{figure}[!b]
\epsfxsize=7.58cm
\centerline{\epsfbox{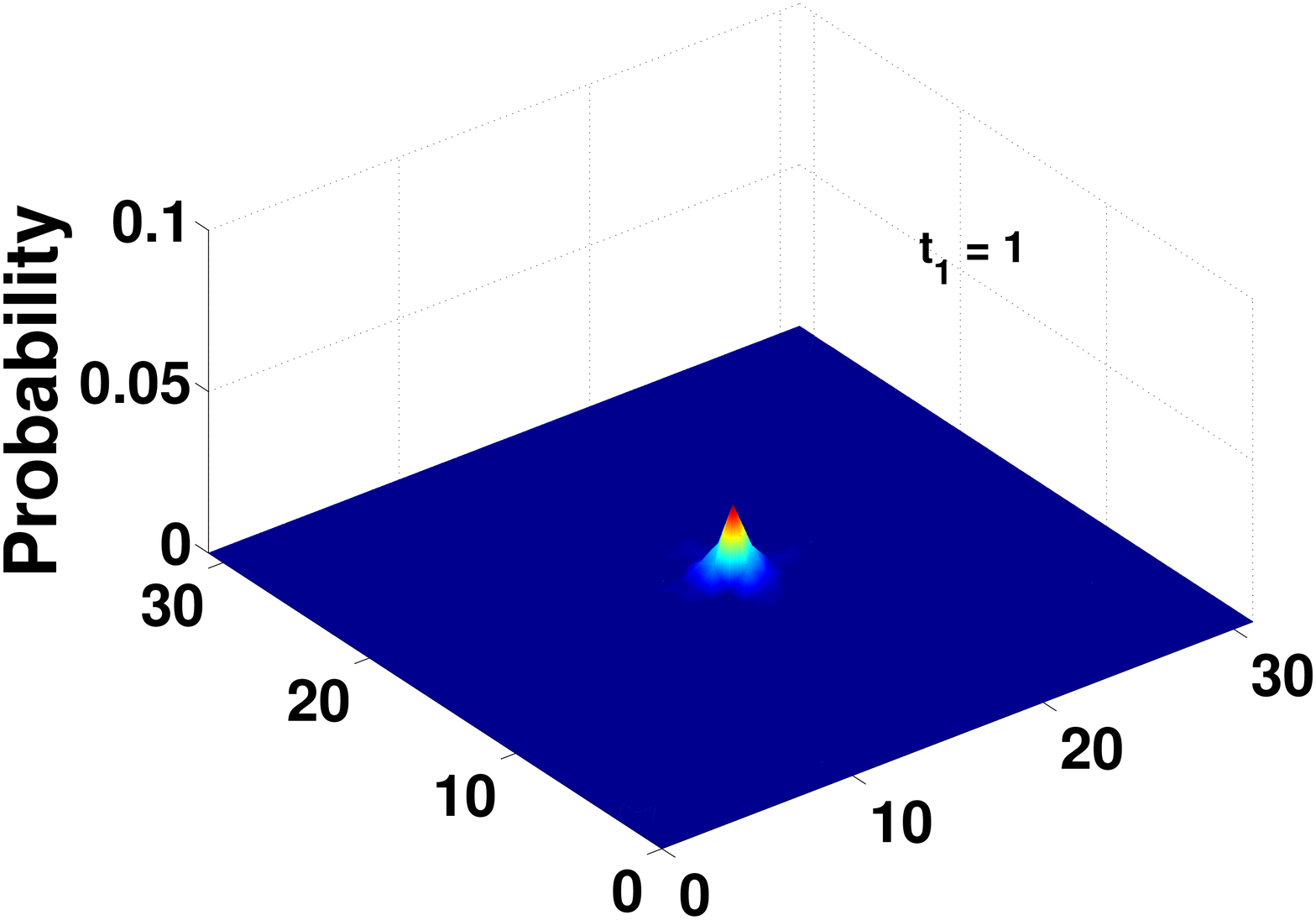}}
\epsfxsize=7.58cm
\centerline{\epsfbox{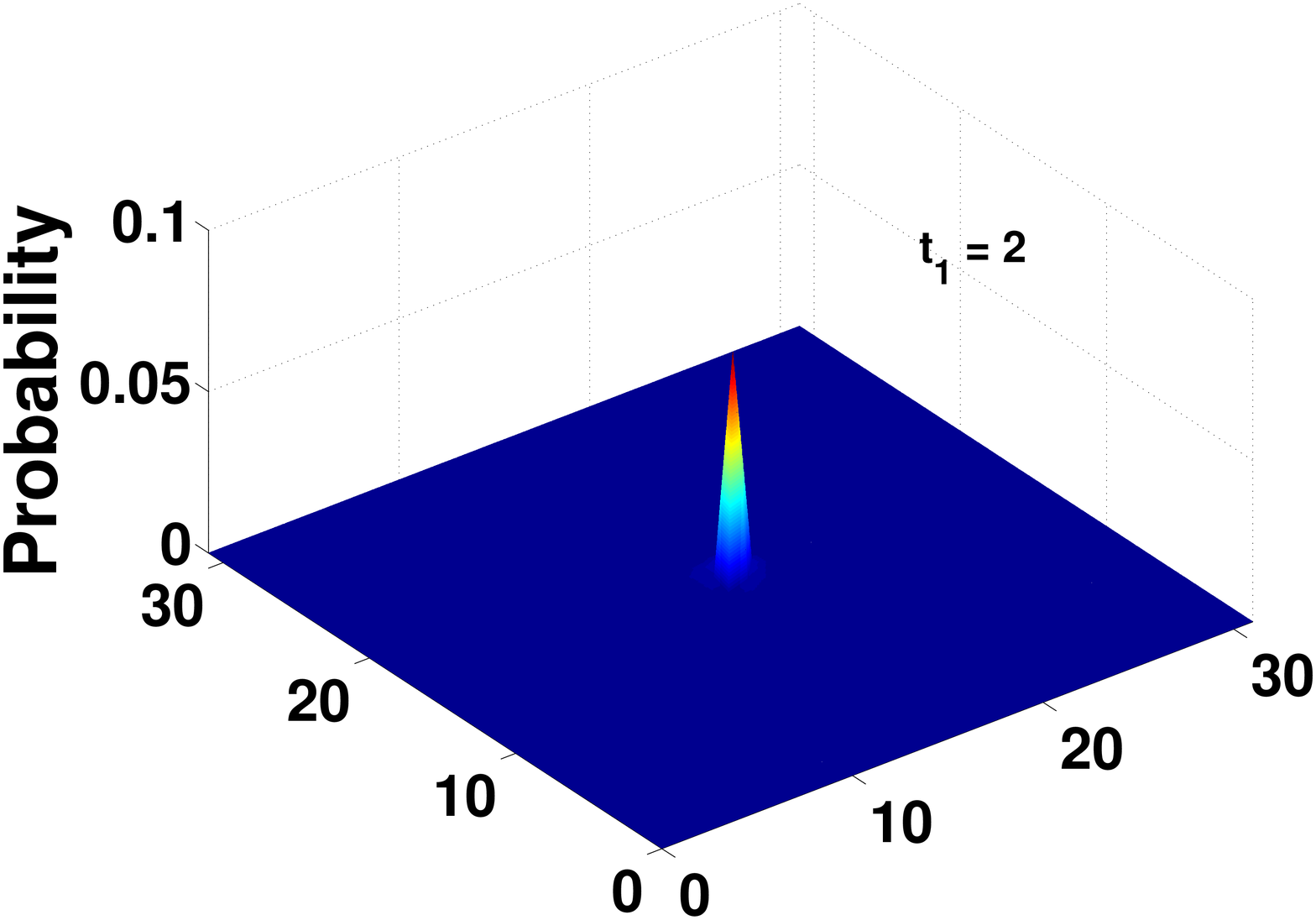}}
\epsfxsize=7.58cm
\centerline{\epsfbox{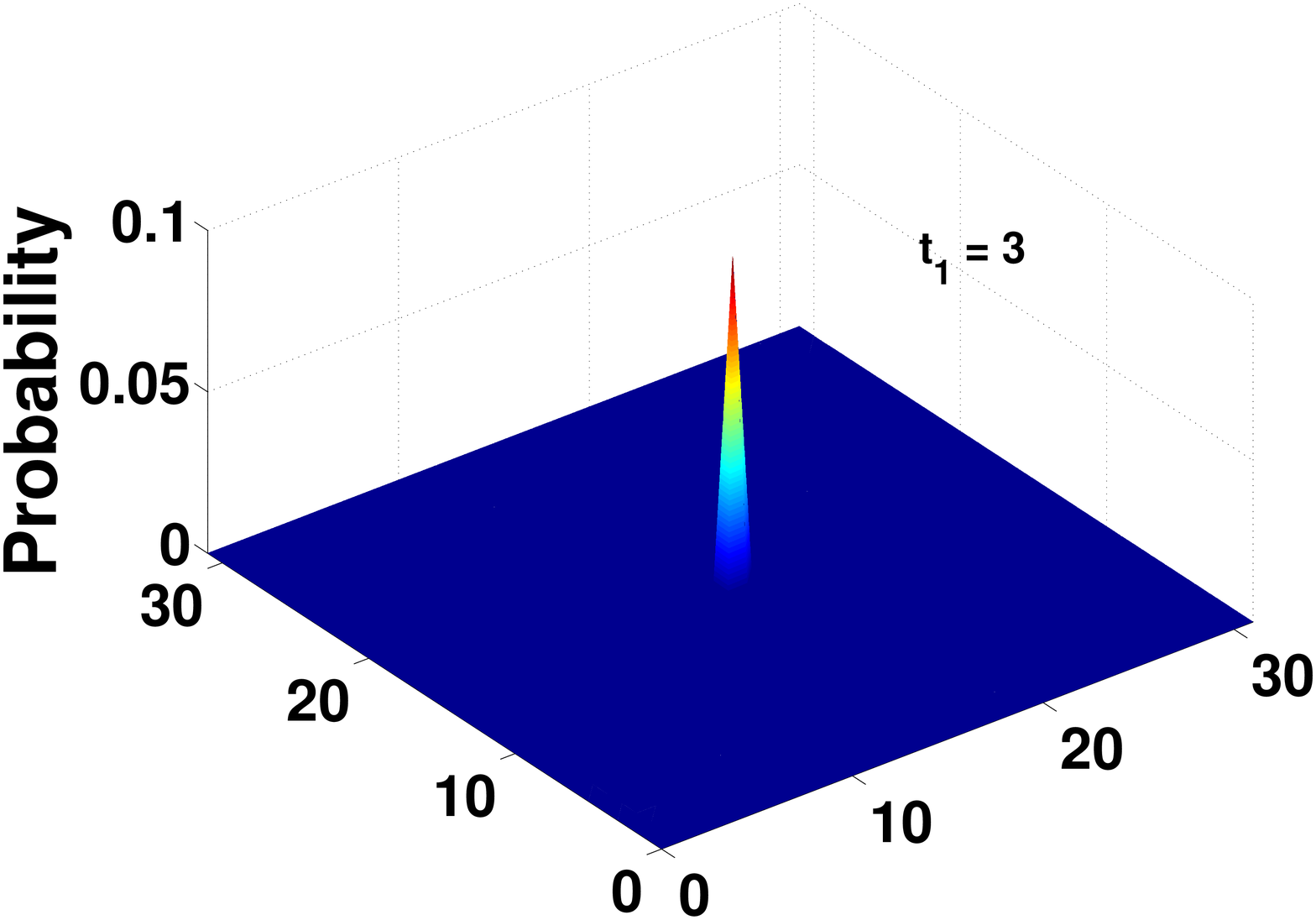}}
\epsfxsize=7.58cm
\centerline{\epsfbox{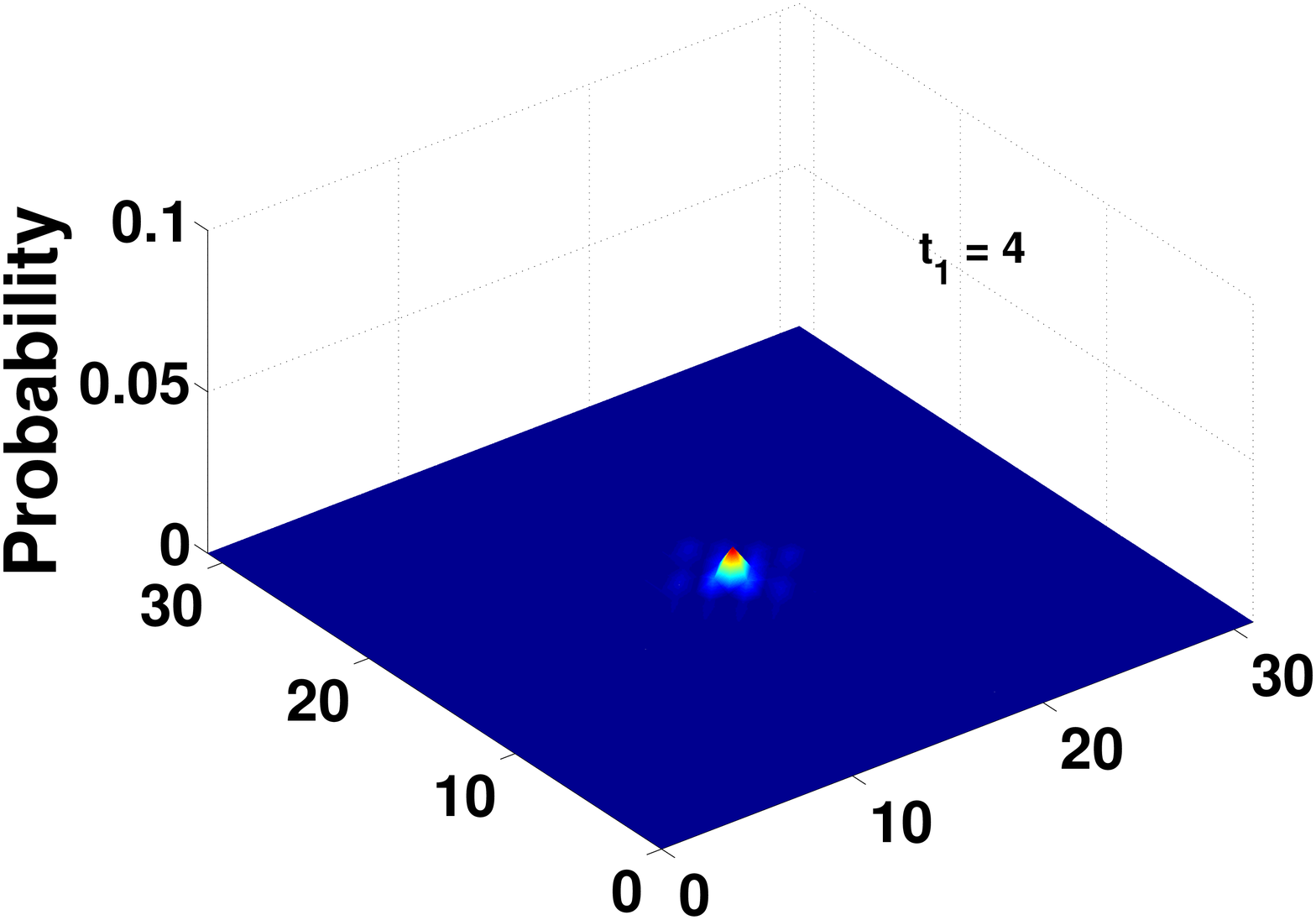}}
\caption{(Color online) Probability distribution for the spatial search
problem on a $32^3$ lattice, at the instance when the probability at the
marked vertex $(16,16,16)$ attains its peak value. The displayed results
are for the $z=16$ slice. $s=1/\sqrt{2}$ and $t_1$ increases from $1$ (top)
to $4$ (bottom).}
\end{figure}

Combining our observations in this particular example, we find that the
best results---the shortest period, the largest peak probability and the
sharpest peak---clearly correspond to $t_1=3$. The results improve as $t_1$
increases from $1$ to $3$, and rapidly deteriorate thereafter. The important
fact that all optimal features appear simultaneously at $t_1=3$ simplifies
tuning of the parameters.

Next we scanned through values of $s$, for different values of $t_1$
and in different dimensions, to determine the optimal values at which the
peak probability $P$ at the marked vertex is the largest and the time
$t_2$ required to reach it is the smallest. A typical situation, with
$t_1=3$ on a $32^3$ lattice, is shown in Fig.4. We notice that continuous
$P$ has a smoother behaviour than discrete $t_{2}$, and maximisation of
$P$ and minimisation of $t_{2}$ occur at roughly the same value of $s$.
So we simplify matters, and select maximisation of $P$ as our optimality
condition.

\begin{figure}[b]
\epsfxsize=9cm
\centerline{\epsfbox{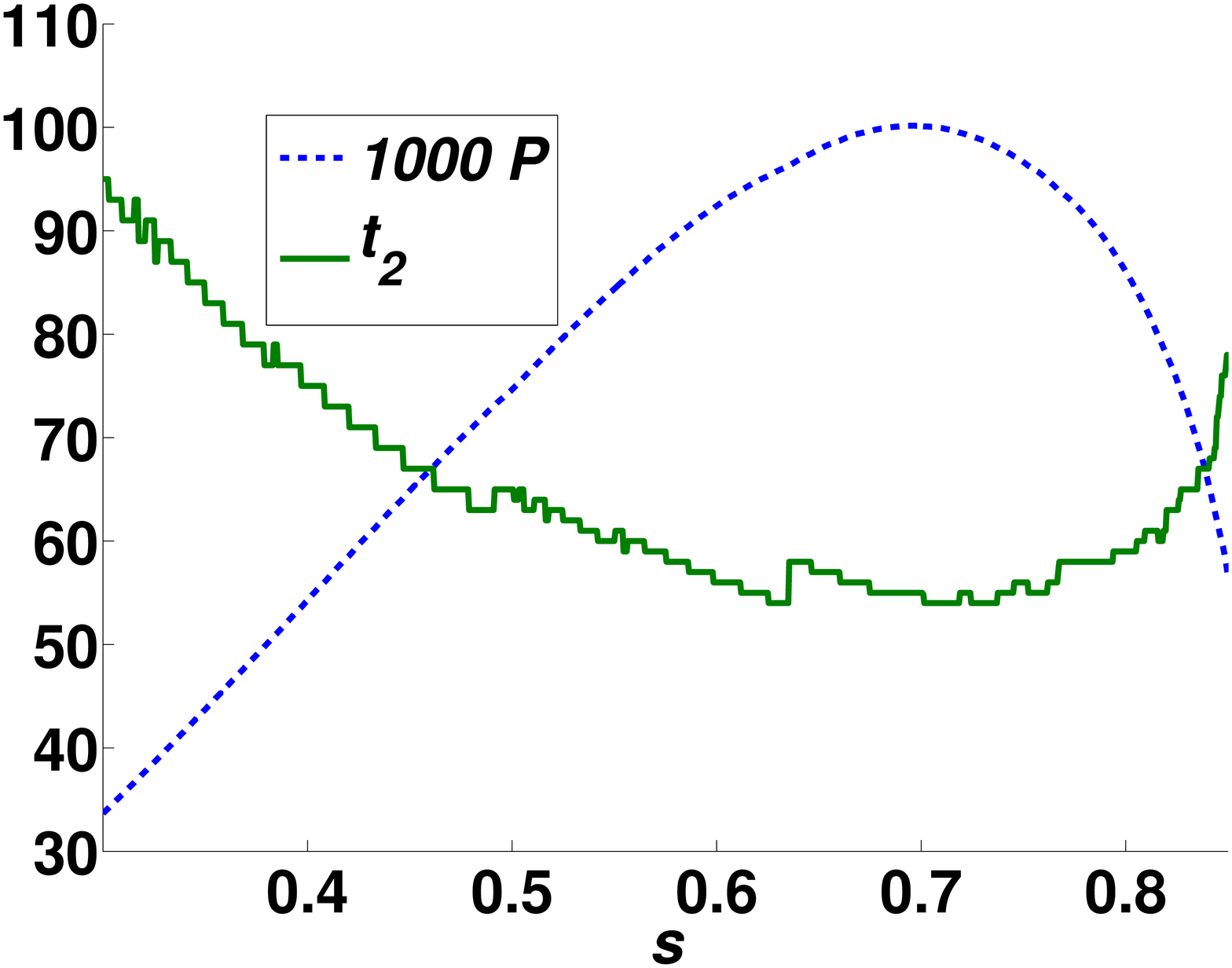}}
\caption{(Color online) Optimisation of $s$ for $t_1=3$ on a $32^3$ lattice,
using peak probability $P$ as well as oracle queries $t_2$.}
\end{figure}

A comparison of our scans, for $t_1=3$ and different number of dimensions,
is presented in Fig.5. We find that at fixed $t_1$, the shape of the $P$
vs. $s$ curve is almost dimension independent, and the optimal value of
$s$ is approximately the same for all dimensions. We also observe that
for fixed $s$, $P$ roughly scales as $2^{-d}$. In the staggered fermion
implementation of the Dirac operator, different vertices of an elementary
hypercube correspond to different degrees of freedom~\cite{staggered}.
So our observation suggests that to a large extent only the degree of
freedom corresponding to the marked vertex evolves during the search,
while other degrees of freedom remain spectators. Moreover, the feature
that the maximum value of $P$ approaches $2^{-d}$ with increasing $d$
implies that the efficiency of our algorithm increases with $d$.
 
\begin{table}[t]
\begin{center}
\caption{Results of optimal parameter determination for different $t_1$ and
in different dimensions.}
\begin{tabular}{|c|c|c|c|c|c|c||c|c|c|}\hline
\multirow{2}{*}{$d$} & \multirow{2}{*}{$L$} & \multirow{2}{*}{$t_1$} & \multicolumn{4}{c||}{Spatial Search} & \multicolumn{3}{c|}{Walk} \\ \cline{4-10}
  &    &   & $s$ & $t_2$ & $P$ & $\theta$ & $s$ & $A(t_1)_{\rm min}$ & $\theta$ \\ \hline
\multirow{5}{*}3 & \multirow{5}{*}{32} & 2 & 0.9507 & 59 & 0.0942  & 3.551 & 0.9258 & \(-\)0.7143 & 3.346\\ 
  &    & 3 & 0.7015 & 55 & 0.1001  & 3.299 & 0.6737 & \(-\)0.7618 & 3.136 \\
  &    & 4 & 0.5363 & 55 & 0.1016  & 3.202 & 0.5194 & \(-\)0.7748 & 3.089 \\
  &    & 8 & 0.2755 & 54 & 0.1027  & 3.158 & 0.2665 & \(-\)0.7860 & 3.052 \\
  &    & 20 & 0.1114 & 54 & 0.1030 & 3.157 & 0.1074 & \(-\)0.7890 & 3.044 \\ \hline
\multirow{5}{*}4 & \multirow{5}{*}{16} & 2 & 0.9541 & 54 & 0.0528  & 3.583 & 0.9428 & \(-\)0.7778 & 3.482 \\
  &    & 3 & 0.6986 & 54 & 0.0548  & 3.281 & 0.6827 & \(-\)0.8190 & 3.188 \\
  &    & 4 & 0.5411 & 53 & 0.0553  & 3.234 & 0.5257 & \(-\)0.8300 & 3.131 \\
  &    & 8 & 0.2771 & 52 & 0.0558  & 3.177 & 0.2694 & \(-\)0.8395 & 3.086 \\
  &    & 20 & 0.1115 & 52 & 0.0559  & 3.160& 0.1084 & \(-\)0.8420 & 3.072 \\ \hline
\multirow{5}{*}5 & \multirow{5}{*}{16} & 2 & 0.9500 & 150 & 0.0276 & 3.545 & 0.9535 & \(-\)0.8182 & 3.577 \\
  &    & 3 & 0.6920 & 148 & 0.0284 & 3.242 & 0.6880 & \(-\)0.8541 & 3.219 \\
  &    & 4 & 0.5376 & 147 & 0.0286 & 3.211 & 0.5292 & \(-\)0.8636 & 3.155 \\
  &    & 8 & 0.2726 & 147 & 0.0288 & 3.124 & 0.2710 & \(-\)0.8718 & 3.105 \\
  &    & 20 & 0.1108 & 147 & 0.0288 & 3.140& 0.1092 & \(-\)0.8739 & 3.095 \\ \hline	
\multirow{2}{*}6 & \multirow{2}{*}{8} & 3 & 0.6891 & 51 & 0.0145 & 3.225 & 0.6913 & \(-\)0.8778 & 3.238 \\
  &    & 20 & 0.1106 & 51 & 0.0147 & 3.135 & 0.1094 & \(-\)0.8951 & 3.101 \\ \hline
\multirow{2}{*}7 & \multirow{2}{*}{8} & 3 & 0.6932 & 102 & 0.0073 & 3.250 & 0.6937 & \(-\)0.8949 & 3.252 \\
  &    & 20 & 0.1097 & 102 & 0.0074 & 3.109 & 0.1098 & \(-\)0.9102 & 3.112 \\ \hline
\end{tabular}
\end{center}
\end{table}

\begin{figure}[b]
\epsfxsize=9cm
\centerline{\epsfbox{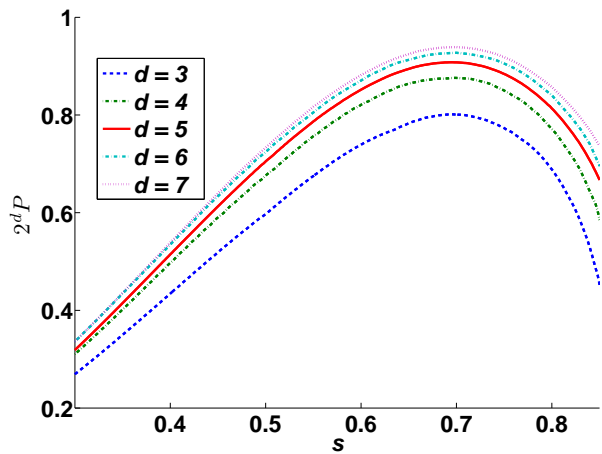}}
\caption{(Color online) Optimisation of $s$ for $t_1=3$ in different
dimensions. Lattice sizes $L=32,16,16,8,8$ were used for $d=3,4,5,6,7$,
respectively.}
\end{figure}

All our quantitative results for optimal values of $s$ are collected in
Table~I. We note that the best performance parameters, $P$ and $t_{2}$,
depend on $d$ but hardly on $t_1$ or $s$. (With increasing $t_1$, there
is a slight increase in $P$ and a small decrease in $t_2$, but that is a
rather tiny improvement.) More interestingly, we find a relation between
the optimal values of $s$ and $t_1$, in terms of the variable
\begin{equation}
\theta = \sqrt{2}t_{1}\sin^{-1}s = t_{1}\tau\sqrt{d/2} ~.
\label{thetadef}
\end{equation}
Since $\sin^{-1}s$ describes the unitary rotation performed by the single
step operators $U_{o(e)}$, and $t_1$ is the number of walk steps, $\theta$
is a measure of the unitary rotation performed by the operator $W^{t_1}$.
The corresponding operator in Grover's optimal algorithm, $G$, is a
reflection providing a phase change of $\pi$. We find that our optimal
$W^{t_1}$ is also close to a reflection, in a sense, with $\theta\approx\pi$.
Unlike the discrete $\pm 1$ eigenvalues of $G$ though, eigenvalues of
$W^{t_1}$ form a distribution. We do not know this distribution in general,
but we know it in the limit $\tau\rightarrow0$, as described at the end of
Subsection~II.A. This limit corresponds to small $s$ and large $t_1$,
and Table~I shows that $\theta$ indeed gets close to $\pi$ in this limit.
To be precise, this normalisation of $\theta$ implies that the optimal
$W^{t_1} \equiv e^{-i\tilde{H}\tau t_1}$ is a reflection operator for
the ``average eigenvalue'' of the associated $\tilde{H}^{2}$. Appearance
of the average eigenvalue in the normalisation indicates that all spatial
modes contribute to the search process with roughly equal strength.

Table~I also lists our results for the variable $A(t_1)$ defined in
Subsection II.A, which characterises how close $W^{t_1}$ is to $G$.
We computed $A(t_1)$, by starting with unit amplitude at the marked vertex,
and then measuring the amplitude at the marked vertex after $t_1$ walk steps.
At the beginning of this subsection we mentioned that checking $A(t_1)$ is
not as elaborate an optimisation test as using the full search algorithm.
Nevertheless, the results corroborate our optimisation of spatial search:
The minimum value of $A(t_{1})$ is reasonably close to $-1$, and occurs at
approximately the same value of $s$ (and hence $\theta$) that provides the
optimal search (the mismatch decreases with increasing $d$). We also observe
that $A(t_1)$ gets closer to $-1$, suggesting that the algorithm becomes more
efficient, with increasing $d$ as well as $t_1$.

Given the strong correlation between the optimal values of $s$ and $t_1$,
the computational cost of the algorithm is minimised by choosing the
smallest $t_1$. For $t_1=1$, Eq.(\ref{thetadef}) does not have a solution
with $\theta\approx\pi$, the optimal value of $s$ is very close to $1$,
and our algorithm does not work well there. Then $t_1=2$ is the choice
that minimises the number of walk steps in the algorithm. Even when the
emphasis is on optimising $P$ and $t_2$, most of the improvement can be
obtained by going from $t_1=2$ to $t_1=3$, and we need not go beyond that.

\section{Simulation Results}

\begin{figure}[b]
\epsfxsize=9cm
\centerline{\epsfbox{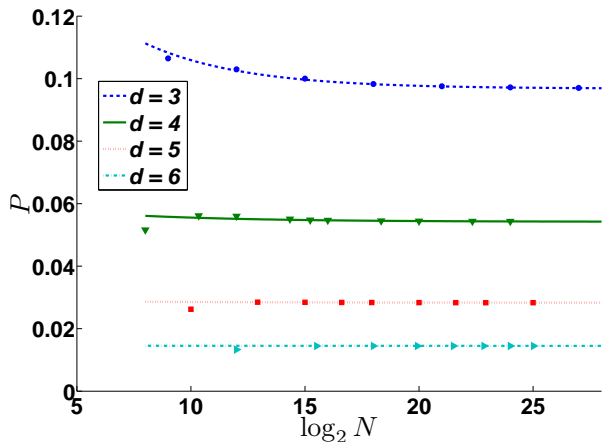}}
\caption{(Color online) Peak probability at the marked vertex as a
function of database size (ranging from $2^8$ to $2^{27}$) in different
dimensions. The points are the data from the simulations with $t_1=3$,
and the curves are the fits $P=a_1+(b_1/L)$.}
\end{figure}

\begin{figure}[b]
\epsfxsize=9cm
\centerline{\epsfbox{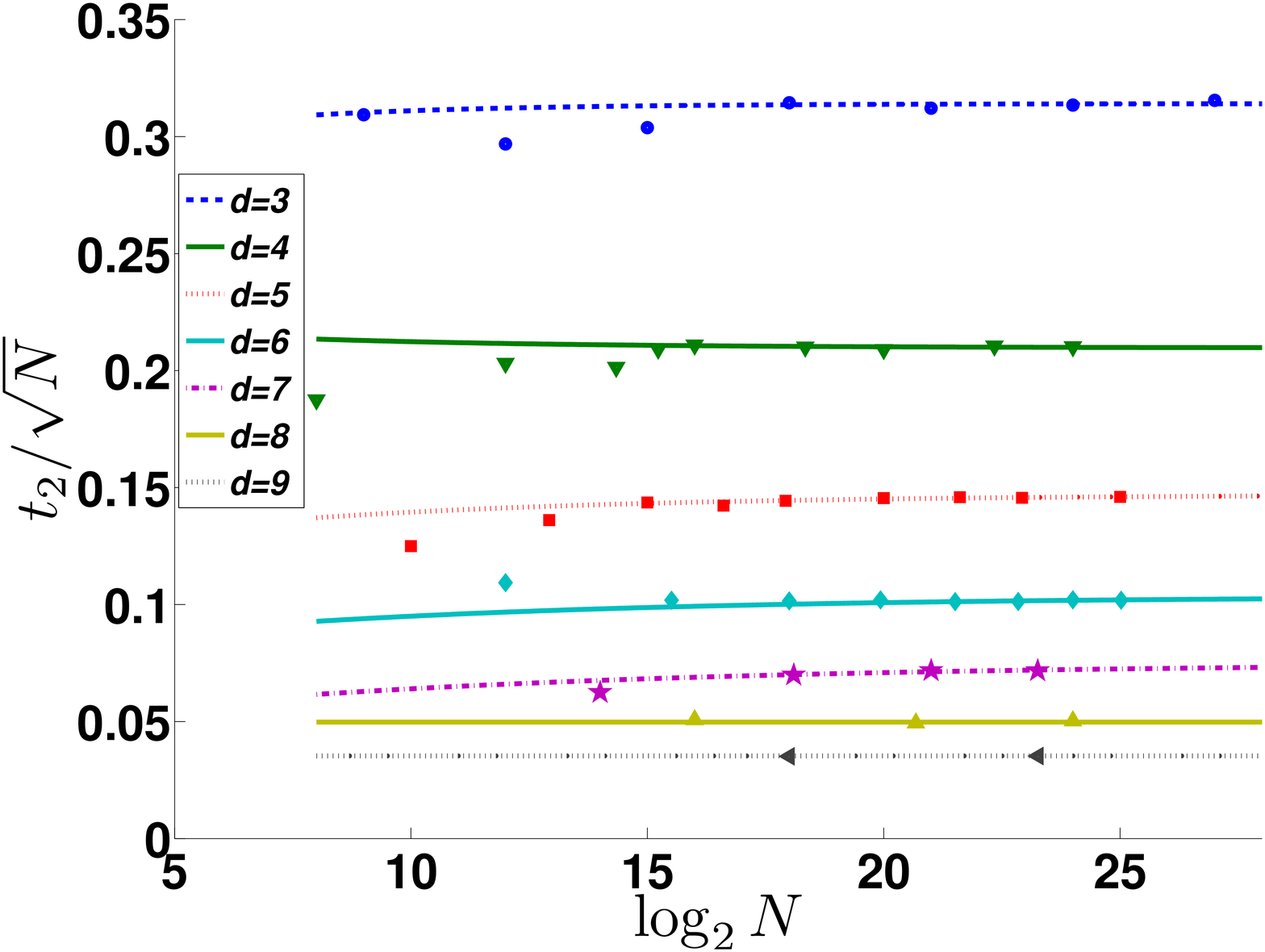}}
\caption{(Color online) Number of search oracle queries as a function
of database size (ranging from $2^8$ to $2^{27}$) in different dimensions.
The points are the data from the simulations with $t_1=3$, and the curves
are the fits $(t_2/\sqrt{N})=a_2+(b_2/L)$.}
\end{figure}

We carried out numerical simulations of the quantum spatial search problem,
with a single marked vertex and lattice dimension ranging from $3$ to $9$.
The matrices $U_{o(e)}$ are real with our conventions, and that was
convenient for simulations. The optimal values of $s$-$t_1$ pairs depend
a little on the lattice size and dimension. We are only interested in the
asymptotic behaviour of the algorithm, however, so we used the same
parameter values for all lattice sizes and dimensions: $s=0.9539$ for
$t_1=2$, $s=1/\sqrt{2}$ for $t_1=3$ and $s=0.5410$ for $t_1=4$.

To extract the scaling properties of the algorithm, we studied the behaviour
of $P$ and $t_2$, as a function of $L$ and $d$. We found that the finite
size effects in our data, as $L\rightarrow\infty$ at fixed $d$, are well
described by series in inverse powers of $L$. In Figs.6-7, we show some
of our results with the fitting functions:
\begin{equation}
P = a_1 + {b_1 \over L} ~,~~
{t_2 \over \sqrt{N}} = a_2 + {b_2 \over L} ~.
\end{equation}
All the fit parameters are listed in Table~II, where error refers to
the r.m.s. deviation of the data from the fit. We do not have sufficient
data for $d=8$ and $d=9$, to make accurate fits or to see the asymptotic
behaviour. Our analysis, therefore, relies on the patterns seen for $d=3$
to $d=7$, while we use $d=8$ and $d=9$ values only for consistency checks.

\begin{table*}[t]
\begin{center}
\caption{Fit parameters for peak probability $P$ and search oracle queries
$t_2/\sqrt{N}$ vs. $L$.}
\begin{tabular}{|c|c|c|c|c|c|c||c|c|c||c|}\hline
$s$ & $t_1$ & $d$ & $L$ & $a_1$ & $b_1$ & Error & $a_2$ & $b_2$ & Error & $a_2/\sqrt{a_1}$ \\ \hline
\multirow{7}{*}{0.9539} & \multirow{7}{*} 2 & 3 & 64,128,256,512 & 0.0911 & \phantom{\(-\)}0.0964 & 2.43$\times$$10^{-5}$
& 0.3237 & \phantom{\(-\)}0.1440  & 4.57$\times$$10^{-4}$ & 1.072\\ \cline{3-11}
    & & 4 & 16,32,48,64 & 0.0522 & \phantom{\(-\)}0.0100 & 3.02$\times$$10^{-5}$
& 0.2167 & \(-\)0.0832  & 1.05$\times$$10^{-3}$ & 0.948\\ \cline{3-11}
    & & 5 & 12,16,24,32 & 0.0275 & \phantom{\(-\)}0.0009 & 1.70$\times$$10^{-6}$
& 0.1486 & \(-\)0.0243  & 2.39$\times$$10^{-4}$ & 0.896\\ \cline{3-11}
    & & 6 & 12,14,16,18 & 0.0141 & \phantom{\(-\)}0.0001 & 2.31$\times$$10^{-7}$
& 0.1043 & \(-\)0.0208  & 1.78$\times$$10^{-4}$ & 0.878\\ \cline{3-11}
    & & 7 & 6,8,10      & 0.0072 & \(-\)0.0004 & 1.40$\times$$10^{-6}$
& 0.0757 & \(-\)0.0338  & 2.47$\times$$10^{-4}$ & 0.892\\ \cline{3-11}
    & & 8 & 6,8         & 0.0036 & ---         & 3.50$\times$$10^{-6}$
& 0.0509 & ---      & 7.23$\times$$10^{-5}$ & 0.848\\ \cline{3-11}
    & & 9 & 6           & 0.0018 & ---         & --- 
& 0.0356 & ---      & ---                   & 0.839\\ \hline

\multirow{7}{*}{1/$\sqrt{2}$} & \multirow{7}{*} 3 & 3 & 64,128,256,512 & 0.0968 & \phantom{\(-\)}0.0920 & 2.73$\times$$10^{-6}$
& 0.3141 & \(-0.0306\)  & 1.23$\times$$10^{-3}$ & 1.010\\ \cline{3-11}
    & & 4 & 16,32,48,64 & 0.0542 & \phantom{\(-\)}0.0076 & 1.46$\times$$10^{-5}$
& 0.2097 & \phantom{\(-\)}0.0151  & 6.71$\times$$10^{-4}$ & 0.901\\ \cline{3-11}
    & & 5 & 12,16,24,32 & 0.0283 & \phantom{\(-\)}0.0010 & 4.66$\times$$10^{-6}$
& 0.1470 & \(-\)0.0300  & 2.12$\times$$10^{-4}$ & 0.874\\ \cline{3-11}
    & & 6 & 12,14,16,18 & 0.0145 & \phantom{\(-\)}0.0001 & 5.79$\times$$10^{-7}$
& 0.1035 & \(-\)0.0269  & 1.88$\times$$10^{-4}$ & 0.860\\ \cline{3-11}
    & & 7 & 6,8,10      & 0.0074 & \(-\)0.0003 & 1.46$\times$$10^{-6}$
& 0.0750 & \(-\)0.0296  & 3.39$\times$$10^{-4}$ & 0.872\\ \cline{3-11}
    & & 8 & 6,8         & 0.0037 & ---         & 3.00$\times$$10^{-6}$
& 0.0498 &  ---     & 4.55$\times$$10^{-5}$ & 0.819\\ \cline{3-11}
    & & 9 & 6           & 0.0019 & ---         & ---
& 0.0353 &  ---     & ---                   & 0.810\\ \hline

\multirow{7}{*}{0.5410} & \multirow{7}{*} 4 & 3 & 64,128,256,512 & 0.0984 & \phantom{\(-\)}0.0936 & 1.84$\times$$10^{-5}$
& 0.3123 & \(-\)0.1239  & 8.46$\times$$10^{-4}$ & 0.996\\ \cline{3-11}
    & & 4 & 16,32,48,64 & 0.0548 & \phantom{\(-\)}0.0087 & 2.24$\times$$10^{-5}$
& 0.2103 & \(-\)0.0500  & 3.57$\times$$10^{-4}$ & 0.898\\ \cline{3-11}
    & & 5 & 12,16,24,32 & 0.0285 & \phantom{\(-\)}0.0013 & 8.05$\times$$10^{-6}$
& 0.1455 & \(-\)0.0142  & 2.25$\times$$10^{-4}$ & 0.862\\ \cline{3-11}
    & & 6 & 12,14,16,18 & 0.0146 & \phantom{\(-\)}0.0002 & 7.67$\times$$10^{-7}$
& 0.1015 & \phantom{\(-\)}0.0043  & 6.28$\times$$10^{-5}$ & 0.840\\ \cline{3-11}
    & & 7 & 6,8,10      & 0.0074 & \(-\)0.0003 & 2.86$\times$$10^{-6}$
& 0.0733 & \(-\)0.0194  & 2.01$\times$$10^{-4}$ & 0.852\\ \cline{3-11}
    & & 8 & 6,8         & 0.0037 & ---         & 4.50$\times$$10^{-6}$
& 0.0505 &  ---     & 4.39$\times$$10^{-4}$ & 0.830\\ \cline{3-11}
    & & 9 & 6           & 0.0019 & ---         & ---
& 0.0353 &  ---     & ---                   & 0.810\\ \hline
\end{tabular}
\end{center}
\end{table*}

\begin{figure}[b]
\epsfxsize=9cm
\centerline{\epsfbox{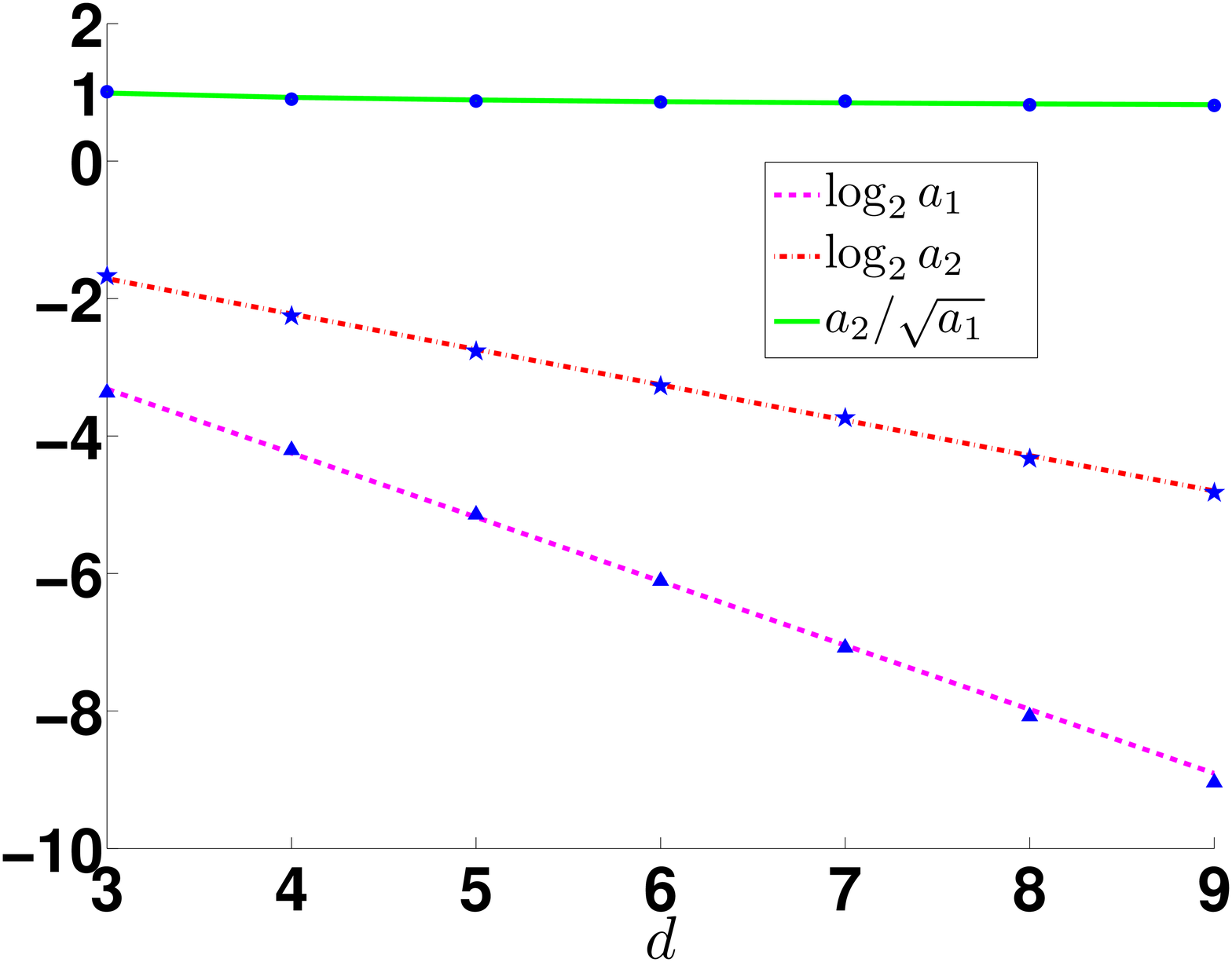}}
\caption{(Color online) Behaviour of the asymptotic fit parameters as a
function of dimension. The data points are from Table~II, with $t_1=3$,
and the curves are the fits described in Eqs.(\ref{fitd1},\ref{fitd2}).}
\end{figure}

We observe the following features:\\
(1) There is not much difference among the $a_1$ and $a_2$ values for
different optimised $s$-$t_1$ pairs. This confirms our earlier inference
that implementing our algorithm for $t_1=2$ minimises the running cost.\\
(2) $P$ approaches a constant as $L\rightarrow\infty$, with $a_1$ decreasing
approximately as $2^{-d}$ with increasing $d$. As mentioned before, this is
a consequence of the fact that in the staggered fermion implementation of
the Dirac operator, only the degree of freedom corresponding to the location
of the marked vertex on the elementary hypercube participates in the search
process. Since the number of vertices in an elementary hypercube is $2^d$,
$P$ is upper bounded by $2^{-d}$. Our algorithm achieves that bound up to
a constant multiple close to $1$.\\
(3) $b_1$ decreases toward zero with increasing $d$, which is also true
for $|b_2|$ to a large extent. Thus the smallest $d=3$ has the largest
finite size corrections to the asymptotic $L\rightarrow\infty$ behaviour.
This suggests that as the number of neighbours of a vertex increase,
the finite size of the lattice becomes less and less relevant.\\
(4) $t_2$ is proportional to $\sqrt{N}$ as $L\rightarrow\infty$, with $a_2$
decreasing approximately as $2^{-d/2}$ with increasing $d$. Thus $t_2$ scales
with the database size as $(L/2)^{d/2}=\sqrt{N/2^d}$. This is in agreement
with the fact that, with only one degree of freedom per elementary hypercube
participating in the search process, the effective number of vertices being
searched is $N/2^d$.\\
(5) In the combination $t_2/\sqrt{P}$, describing the effective number of
oracle queries for our algorithm, the factors of $2^d$ arising from the
number of vertices in an elementary hypercube fully cancel. That makes
$t_2/\sqrt{P}$ scale as $\sqrt{N}$, and $a_2/\sqrt{a_1}$ describes the
complexity scaling of our algorithm. As listed in the rightmost column of
Table~II, $a_2/\sqrt{a_1}$ shows a gradual decrease with increasing $d$,
and our results are just above the corresponding value $\pi/4$ for
Grover's optimal algorithm. More explicitly, our result for $d=3$ is about
25\% above $\pi/4$, decreasing to about 10\% above $\pi/4$ for $d=7$.
These values imply that our algorithm improves in efficiency with increasing
$d$, and is not too far from the optimal behaviour even for the smallest
dimension $d=3$.

To analyse the dimension dependence of our algorithm in more detail,
we fit the values in Table~II to the functional forms,
\begin{equation}
\log_2 a_1 = c_1 + d_1 d ~,~~
\log_2 a_2 = c_2 + d_2 d ~,~~
\label{fitd1}
\end{equation}
\begin{equation}
{a_2 \over \sqrt{a_1}} = c_3 + {d_3 \over d} ~.
\label{fitd2}
\end{equation}
The results for $t_1=3$ are displayed in Fig.8, showing that the fits work
reasonably well.

\begin{table*}[t]
\begin{center}
\caption{Fit parameters for $\log_2 a_1$, $\log_2 a_2$ and $a_2/\sqrt{a_1}$
vs. $d$.}
\begin{tabular}{|c|c|c|c|c|c||c|c|c||c|c|c|}\hline
$s$ & $t_1$ & Fitted $d$ & $c_1$ & $d_1$ & Error & $c_2$ & $d_2$ & Error & $c_3$ & $d_3$ & Error \\ \hline
0.9539       & 2 & \multirow{3}{*}{3,4,5,6,7,8} & \(-\)0.553 & \(-\)0.938 & 4.77$\times 10^{-2}$
       & \(-\)0.085 & \(-\)0.526 & 2.53$\times 10^{-2}$ & 0.721 & 0.995 & 1.71$\times 10^{-2}$ \\
$1/\sqrt{2}$ & 3 &                              & \(-\)0.458 & \(-\)0.947 & 4.48$\times 10^{-2}$
       & \(-\)0.138 & \(-\)0.521 & 2.57$\times 10^{-2}$ & 0.729 & 0.791 & 1.63$\times 10^{-2}$ \\
0.5410	     & 4 &                              & \(-\)0.421 & \(-\)0.951 & 3.78$\times 10^{-2}$
       & \(-\)0.151 & \(-\)0.521 & 2.09$\times 10^{-2}$ & 0.725 & 0.761 & 1.32$\times 10^{-2}$ \\ \hline
\end{tabular}
\end{center}
\end{table*}

\begin{figure}[b]
\epsfxsize=9cm
\centerline{\epsfbox{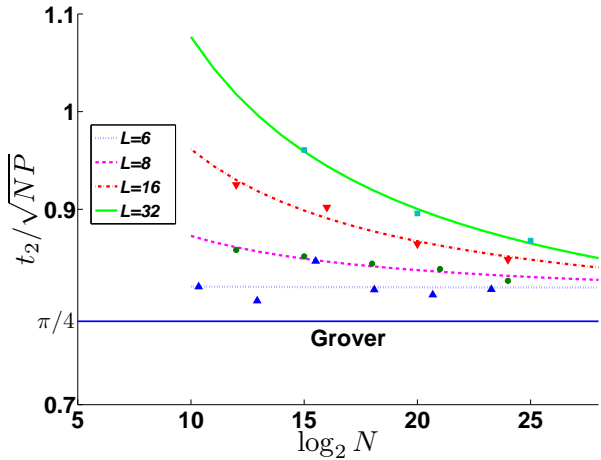}}
\caption{(Color online) $t_2/\sqrt{NP}$ as a function of the database size
for different lattice sizes. The points are the data for $t_1=3$, and the
curves are the fits $(t_2/\sqrt{NP})=a_l+(b_l/d)$. Also shown is the
limiting value corresponding to Grover's optimal algorithm.}
\end{figure}

The fit parameters for the three different optimised $s$-$t_1$ pairs are
listed in Table~III. They support our earlier scaling observations: $d_1$
close to $-1$ means that $P$ scales roughly as $2^{-d}$, and $d_2$ close
to $-0.5$ means that $t_2$ scales roughly as $\sqrt{N/2^d}$.
Interpreting Grover's algorithm as the $d\rightarrow\infty$ limit of
the spatial search algorithm, we expect $c_3$ to be close to $\pi/4$.
It turns out to be somewhat smaller, but is not very accurately determined
due to sizable correction from $d_3$ for the values of $d$ that we have.
We believe that simulations for larger lattice sizes and higher dimensions
can remedy this problem.

To look at our data in a different manner, we analyse the scaling behaviour
of $t_2/\sqrt{NP}$ as a function of the database size. In Fig.9, we plot a
subset of the data, illustrating how $t_2/\sqrt{NP}$ behaves for fixed
lattice size $L$ when the database size is changed by changing $d$.
The approach to the asymptotic behaviour is particularly clear in this
form, with $t_2/\sqrt{NP}$ decreasing toward $\pi/4$ as the database
size increases. It is also obvious that, for fixed $N$, our algorithm
improves in efficiency with decreasing $L$ (or equivalently with increasing
$d$). This is a consequence of the increase in the number of neighbours of
a vertex with increasing $d$. An important implication is that, given a
database of size $N$, it is best to implement the algorithm using the
smallest $L$ (and hence the largest $d$) available. For our staggered
fermion implementation of the Dirac operator, $L$ has to be even, and
the smallest $L$ possible is $4$. Because of small values of $d$ that
we have worked with, and discrete nature of $t_2$, our data for $L=4$
do not show smooth asymptotic behaviour. We therefore omitted $L=4$ from
our analysis, and our best results are for $L=6$.

We can fit the data reasonably well as
\begin{equation}
{t_2 \over \sqrt{NP}} = a_l +{b_l \over d} ~.
\end{equation}
The fit parameters for the three different optimised $s$-$t_1$ pairs are
listed in Table~IV. We note that $a_l$ is close to $\pi/4$ when data for
$d>6$ are available, but it cannot be accurately determined using data for
small $d$ only due to sizable corrections from $b_l$.

\begin{table}[b]
\begin{center}
\caption{Fit parameters for $t_{2}/\sqrt{NP}$ vs. $d$ at fixed $L$.}
\begin{tabular}{|c|c|c|c|c|c|c|}\hline
$s$ & $t_1$ & $L$ & Fitted $d$ & $a_l$ & $b_l$ & Error \\ \hline
\multirow{4}{*}{0.9539} & \multirow{4}{*}{2} & 6 & {5,6,7,8,9} & 0.766 & \phantom{$-$}0.551 & 1.06$\times 10^{-2}$ \\
  &  &  8 & {4,5,6,7,8} & 0.805 & \phantom{$-$}0.293 & 5.31$\times 10^{-4}$ \\
  &  & 16 & {3,4,5,6}   & 0.663 & \phantom{$-$}1.132 & 1.73$\times 10^{-2}$ \\
  &  & 32 & {3,4,5}     & 0.625 & \phantom{$-$}1.304 & 5.77$\times 10^{-3}$ \\ \hline
\multirow{4}{*}{$1/\sqrt{2}$} & \multirow{4}{*}{3} & 6 & {4,5,6,7,8,9} & 0.819 & \phantom{$-$}0.006 & 1.28$\times 10^{-2}$ \\
  &  &  8 & {4,5,6,7,8} & 0.803 & \phantom{$-$}0.234 & 3.28$\times 10^{-3}$ \\
  &  & 16 & {3,4,5,6}   & 0.773 & \phantom{$-$}0.471 & 6.46$\times 10^{-3}$ \\
  &  & 32 & {3,4,5}     & 0.724 & \phantom{$-$}0.705 & 3.23$\times 10^{-3}$ \\ \hline
\multirow{4}{*}{0.5410} & \multirow{4}{*}{4} & 6 & {4,5,6,7,8,9} & 0.837 & $-$0.092 & 1.35$\times 10^{-2}$ \\
  &  &  8 & {4,5,6,7,8} & 0.791 & \phantom{$-$}0.272 & 3.79$\times 10^{-3}$ \\
  &  & 16 & {3,4,5,6}   & 0.767 & \phantom{$-$}0.450 & 1.29$\times 10^{-3}$ \\
  &  & 32 & {3,4,5}     & 0.711 & \phantom{$-$}0.726 & 1.13$\times 10^{-3}$ \\ \hline
\end{tabular}
\end{center}
\end{table}

\section{Results for Multiple\break Marked Vertices}

We also carried out a few simulations of spatial search with more than one
marked vertex. The change from the single marked vertex algorithm is that
the search oracle now flips the sign of amplitudes at several distinct
vertices, say $M$. To see clear patterns in the results, we needed to
keep the marked vertices well separated and avoid interference effects.

\begin{table*}[t]
\begin{center}
\caption{Spatial search results for multiple marked vertices in $d=3$.}
\begin{tabular}{|c|c|c|c|c|c|}\hline
$s$ & $t_1$ & $M$ & Marked vertices & $P$ & $t_2$ \\ \hline
\multirow{6} {*}{$1/\sqrt{2}$} & \multirow{6} {*}3 & 1 & (32,32,32) & 0.09829 & 161 \\ \cline{3-6}
  &  &  \multirow{2} {*}2 & (0,32,32), (32,32,32)  & 0.04919, 0.04919  & 112, 112   \\ \cline{4-6}  
  &  &  & (0,32,33), (32,32,32)  & 0.09868, 0.09790 & 161, 161  \\ \cline{3-6}
  &  &  \multirow{3} {*}3 &(0,0,0), (16,16,16), (32,32,32)  & 0.03530, 0.03264, 0.03082  & 94, 92, 92   \\ \cline{4-6} 
  &  &  & (0,0,1), (16,16,16), (32,32,32) & 0.09380, 0.05590, 0.04507  & 161, 117, 109   \\  \cline{4-6}
  &  &  & (0,0,1), (16,16,16), (33,32,32) & 0.09298, 0.09838, 0.10347  & 157, 161, 162   \\  \hline
\end{tabular}
\end{center}
\end{table*}

With the staggered fermion implementation of the Dirac operator, it matters
whether the marked vertices belong to the same corner of the elementary
hypercube or to different corners. So we considered various hypercube
locations for the marked vertices. In Table~V, we compare our illustrative
results for two and three marked vertices with those for a single marked
vertex. The values for peak probability $P$ and corresponding time $t_2$
were obtained with $s=1/\sqrt{2}$ and $t_1=3$ on a $64^3$ lattice.

When the marked vertices belong to the same corner of the elementary
hypercube (e.g. (0,32,32) and (32,32,32)), they involve the same degree
of freedom. In that case, we observe that $P$ decreases by a factor of $M$
and $t_2$ decreases by a factor of $\sqrt{M}$. On the other hand, when the
marked vertices belong to different corners of the elementary hypercube
(e.g. (0,32,33) and (32,32,32)), they involve different degrees of freedom.
In that case, we find that $P$ and $t_2$ are more or less the same as for
the single marked vertex case. This pattern is also followed when some
marked vertices are at the same corner of the elementary hypercube and
some at different corners.

We can infer the following equipartition rule from these observations.
A fixed peak probability (upper bounded by $2^{-d}$) is available for
each corner degree of freedom of the elementary hypercube. Multiple
marked vertices corresponding to the same degree of freedom share this
peak probability equally, while different degrees of freedom evolve
independently of each other. Apart from this caveat, the dependence of
$t_2$ on $M$ for our spatial search algorithm is the same as in the case
of Grover's algorithm, i.e. proportional to $\sqrt{N/M}$.

\section{Summary}

We have formulated the spatial search problem using staggered fermion
discretisation of the Dirac operator on a hypercubic lattice in arbitrary
dimension. Our search strategy, Eq.(\ref{evolsearch}), mimics that of
Grover's algorithm. The locality restriction on the walk operator becomes
less and less relevant as the dimensionality of the space increases, and
we expect to reach the optimal scaling behaviour of Grover's algorithm
as $d\rightarrow\infty$.

We have verified our theoretical expectations by numerical simulations
of the algorithm for $d=3$ to $d=9$. We clearly demonstrate the approach
to the $d\rightarrow\infty$ limit, and find that even for $d=3$ our
algorithm is only $25\%$ less efficient. From a computational cost point
of view, for a fixed database size $N$, our best parameters are $t_1=2$
and the largest $d$ available. With the staggered fermion implementation,
our algorithm works in total Hilbert space dimension $N$. That is a big
reduction from the total Hilbert dimension $2^d N$ required by the
algorithms that use coin or internal degrees of freedom (for the best
$d=\Theta(\ln N)$, $2^d=O(N)$ is substantial), and makes our exercise
worthwhile.


\begin{thebibliography}{99}
\bibitem{spinorsize}{This is in spite of the fact that Dirac spinors
         require $2^{[(d+1)/2]}$ components in $d$ dimensions.
         For example, in $d=3$, a Rubik's cube can have the central
         cube labeled ``o'' and the eight corner cubes labeled ``e''.}
\end{thebibliography}
\end{document}